# Optical Degradation Impact on the Spectral Performance of Photovoltaic Technology

Álvaro F. Solas [a,*], L. Micheli [a], Florencia M. Almonacid [a], Eduardo F. Fernández [a,*]

[a] Centre for Advanced Studies in Energy and Environment (CEAEMA), Electronics and Automation Engineering Department, University of Jaén, Las Lagunillas Campus, Jaén, 23071, Spain

## Abstract

The exponential growth of global capacity along with a reduction in manufacturing costs in the last two decades has caused photovoltaic (PV) energy technology to reach a high maturity level. As a consequence, currently, researchers from all over the world are making great efforts to analyse how different types of degradation impact this technology. This study provides a detailed review of the impact of different optical degradation mechanisms, which mainly affect the transmittance of the top-sheet encapsulant, on the spectral response of the PV modules. The evaluation of the impact on the spectral performance of PV modules is evaluated by considering the variations of the short-circuit current since this is the most widely used parameter to study the spectral impact in outdoors. Some of the most common types of optical degradation affecting the performance of PV modules worldwide, such as discoloration, delamination, aging and soiling have been addressed. Due to the widely documented impact of soiling on the spectral response of modules, this mechanism has been specially highlighted in this study. On the other hand, most of the publications analysed in this review report optical degradation in PV modules with polymeric encapsulant materials. Furthermore, an innovative procedure to quantify the spectral impact of degradation on PV modules is presented. This has been used to analyse the impact of two particular cases of degradation due to soiling and discoloration on the spectral response of different PV technologies.

## Keywords

**List of abbreviations**

| | |
|---|---|
| AM | Air Mass |
| AOD | Aerosol Optical Depth |
| AOI | Angle of Incidence |
| a-Si | Amorphous Silicon |

---

* Corresponding author.
 E-mail addresses: afsolas@ujaen.es (Álvaro F. Solas)
                    eduardo.fernandez@ujaen.es (Eduardo F. Fernández)

| | |
|---|---|
| a-Si:H/c-Si | Hydrogenated Amorphous Silicon |
| ATR | Attenuated Total Reflectance |
| BDRatio | Broadband Degradation Ratio |
| BOS | Balance of System |
| CdTe | Cadmium Telluride |
| CIGS | Copper Indium Gallium Selenide |
| CPV | Concentrator Photovoltaic |
| c-Si | Crystalline Silicon |
| D&D | Discoloration and Delamination |
| DRatio | Degradation Ratio |
| EQE | External Quantum Efficiency |
| EVA | Ethyl Vinyl Acetate |
| G | Irradiance |
| IEC | International Electrotechnical Commission |
| LCOE | Levelised Cost of Electricity |
| LID | Light Induced Degradation |
| NIR | Near-infrared |
| P | Power |
| PV | Photovoltaic |
| m-Si | Monocrystalline Silicon |
| PR | Performance Ratio |
| p-Si | Polycrystalline Silicon |
| PW | Precipitable Water |
| QE | Quantum Efficiency |
| SDRatio | Spectral Degradation Ratio |
| SR | Spectral Response |
| STC | Standard Test Condition |
| USA | United States of America |
| UV | Ultraviolet |
| VIS | Visible |

*Nomenclature*

| c | Speed of light |
|---|---|
| e | Electron charge |
| $E_G$ | Global spectrum |
| h | Planck constant |
| $I_{SC}$ | Short-circuit current |
| $J_{SC}$ | Short-circuit current density |
| $n_e$ | Number of electrons |
| $n_p$ | Number of incident photons |
| $R_S$ | Series resistance |
| λ | Wavelength |
| Φ | Luminous flux |
| τ | Spectral transmittance |

*Measurement Units*

| A /W | Ampere per watt |
|---|---|
| C | Coulomb |
| cm | Centimetre |
| GW | Gigawatt |
| J s | Joule-second |
| kWh /m² | Kilowatt per square metre |
| g /m² | Gram per square metre |
| nm | Nanometre |
| mm | Millimetre |
| m /s | Metre per second |
| mg / m² | Milligram per square metre |
| USD / kWh | United States dollar per kilowatt hour |
| W / (m² nm) | Watt per square metre-nanometre |

# 1. Introduction

In recent decades, the global increase in energy consumption and the growing concern about the harmful effects that the energy generation through fossil fuels means for the environment has led to a great development of renewable energies. The energy obtained through renewable sources reached an estimated 17.9% of the global energy consumption in 2018 and 27.3% of global electricity production in 2019 [1]. The main advantage of these energy sources in comparison with fossil fuels is their null or limited impact over the environment. Due to this fact, numerous investigations related to the various types of renewable energy have been carried out in recent years. The main purpose of these investigations is to improve the efficiency of the systems in order to make them an increasingly feasible solution for the commercial energy market.

Among the different types of renewable energies, photovoltaic (PV) solar energy has experienced the greatest growth in recent years. This can be illustrated by the progress of the PV capacity installed in the last few years, which, at the end of 2019, reached a value (627 GW) more than six times larger than that of 2012 (101 GW). This is due to the reduction in PV manufacturing costs, commonly characterised by the levelised cost of electricity (LCOE) [2]. This index expresses the unitary cost of each kWh of electricity generated by a system over its lifetime. The LCOE of PV has significantly fallen between 2010 and 2019, to USD 0.068/ kWh [3]. This value makes PV energy a competitive alternative to the generation of energy through fossil fuels.

Furthermore, the efficiency of different PV technologies has increased substantially in the last 30 years. For example, the efficiency of a thin-film module (CdTe) has grown from 8% in 1993 to 18.6% in 2019 [4]. Other technologies that have experienced important progress are CIGS and monocrystalline silicon modules as it can be seen in Fig. 1.

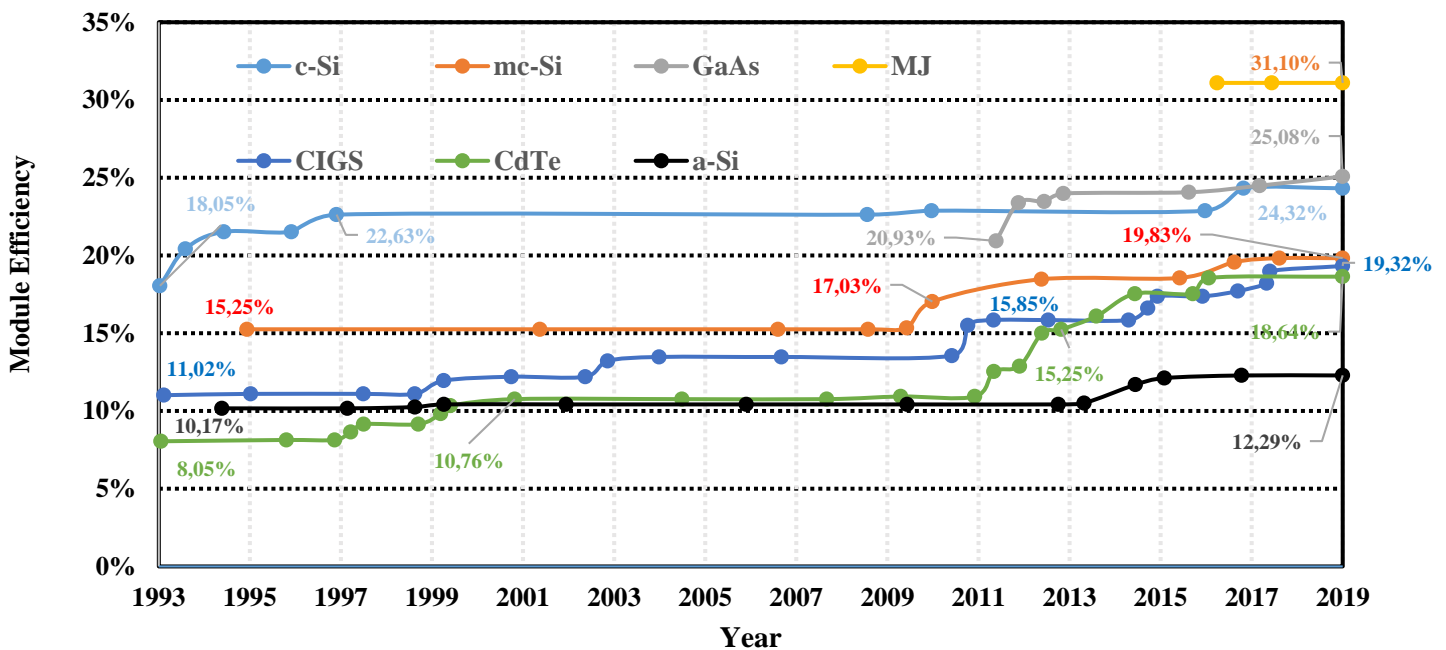

*Fig. 1. Evolution of PV efficiency of different technologies over the last twenty-five years: highest confirmed commercial module results for modules sizes ≥ 800 cm² at AM 1.5G (AM 1.5 D for MJ) [Adapted from 4 and 5].*

Most of the studies performed in the area of photovoltaic solar energy have focused on aspects such as improving the efficiency of PV modules [6–15], reducing costs [16–18], optimizing the design of the balance of system (BOS) [19], and the selection of the optimal orientation and inclination of the modules [20–24] . However, in the last 10 years, the number of publications on the different mechanisms of optical degradation, which affect the behaviour of PV modules [25–27], have significantly increased, as it is shown in Fig. 2.

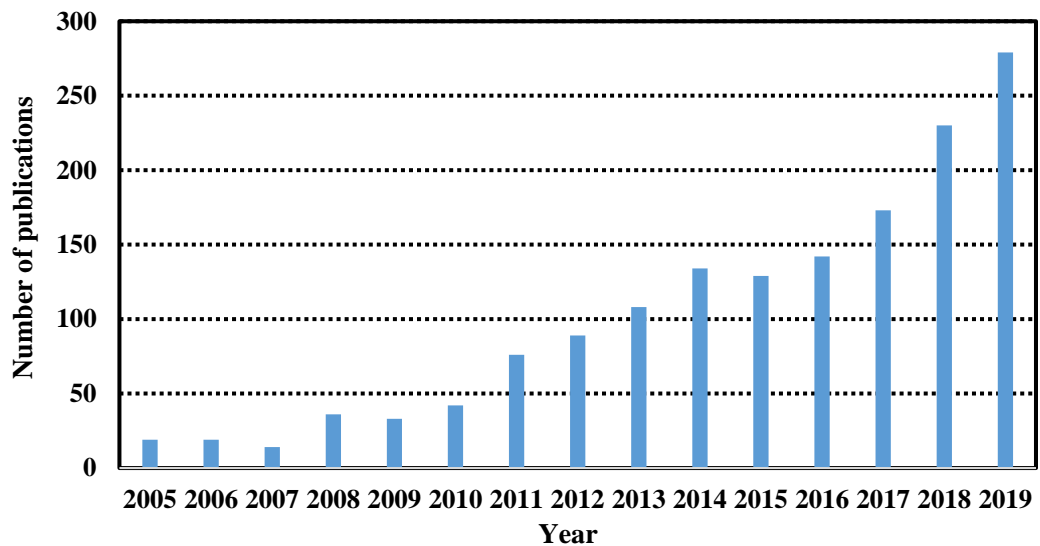

*Fig. 2. Evolution of the number of publications related with degradation in PV modules since 2005. The data have been obtained using Scopus by the search of publications which contain the following words in their title, abstract or keywords: "PV", "degradation" OR "failure" and "module" in July, 2020.*

The importance of the research on the different degradation mechanisms resides in the reduction that these cause to the amount of solar radiation that PV modules can transform into energy, causing both energy and economic losses [28]. For this reason, the analysis of the effects of degradation and the development of preventive techniques should be conducted. The main optical degradation mechanisms are aging, corrosion, encapsulant degradation (discoloration and delamination), atmospheric agents, such as temperature and moisture, and the accumulation of dust and dirt on their surface, known as soiling. All of these factors are expected to affect the spectral behaviour of the PV modules in a greater or minor extent, either for a change in the transmittance of the encapsulant or for a direct alteration of the spectral response of the cells of the module. However, this latter has not been studied in depth yet and the impact on the spectral performance of different PV technologies is still under investigation. For this reason, in our review we recommend to further investigate how degradation impacts the inherent spectral response of PV cells.

Despite this reality, several studies about the effects of the solar spectrum variations on the electrical output of PV devices have been conducted in the last 30 years. Faine et al. 1991 [29] published one of the first works addressing and highlighting this issue. This led to the development of multiple spectral data sets, available on literature [30–32]. The need of the standardization of PV modules characteristics caused the establishment of a reference global irradiance distribution to evaluate PV devices under the called standard test conditions (STC). This reference spectrum was obtained by the analysis of different phenomena and extinction events that happen in the atmosphere. Although the reference spectrum [33], which is indicated in the STC conditions, allows the characterization and standardization of PV devices to be conducted, its occurrence is unlikely in real operation conditions where the knowledge of the electrical output is the main interest [34]. For this reason, researchers are making large efforts to analyse and investigate how PV devices of different technologies are being affected by spectral changes [35–44]. Further, in the last decade, numerous reviews that address the degradation mechanisms mentioned before have been published. Some of them describe the causes and effects of the different mechanisms on PV modules [45–49], whereas others are focused on a particular issue, such as aging [50], light induced degradation (LID) [51], encapsulant degradation [52,53], cell technology [54] and soiling [55–59]. The aim of this study is to provide a comprehensive review on the main optical degradation mechanisms that affect the PV modules, with a special focus on their spectral impacts. The goal is filling the gap concerning the changes on the spectral behaviour due to these mechanisms. The discussed results and conclusions are intended to be a first step towards the achievement of a better understanding of the relationship between the optical degradation mechanisms and the spectral behaviour of PV modules.

In this paper, the main findings that have already been published are summarized and discussed as a way to clarify the spectral impact of different optical degradation mechanisms on PV devices. Most of the presented studies evaluates optical degradation in m-Si and p-Si PV modules, which primarily used a polymeric layer as top-sheet encapsulant [60]. Despite that, optical degradation in other PV technologies, such as CdTe or a-Si, is also mentioned in this work. These two PV technologies make use of glass as top-sheet encapsulant. This fact is expected to clearly affect how different degradation mechanisms impact the optical properties of PV modules' components. Furthermore, the spectral impact of soiling on CPV modules is also reviewed, as under the same conditions, losses due to soiling have been proven to be higher than those of flat PV technologies because of the additional scatter produced by the concentrator optical elements and because of the spectral effects on the current balance of the multijunction cell's subcells. Even if limited information on the spectral impact of some mechanisms has been found, actual spectral data have been used to perform an in-depth analysis, based on theoretical simulations, of some degradation effects, to quantify the spectral losses as a result of the variations of some of the most relevant atmospheric parameters, affecting the irradiance spectrum, i.e. air mass, aerosol optical depth and precipitable water. The purpose of this analysis is to generate a reference approach for the evaluation of the spectral impact of optical degradation mechanisms on PV devices.

The study is structured as follows: Section 2 introduces the fundamentals and the equations which allow the evaluation of the spectral impact on the response of PV devices due to different optical degradation mechanisms; Section 3 presents the review, description and analysis of the mechanisms presented in multiple publications; Section 4 analyses the spectral impact of some of the mechanisms described in Section 3 as a function of the most important atmospheric parameters; finally, Section 5 recapitulates the main conclusions found during the development of this study.

## 2. Theoretical Background

The lack of previous studies related with the spectral impact of degradation mechanisms on PV modules has promoted the development of a novel procedure and methodology to analyse this issue, which is based on the study performed by Fernandez et al. [61]. In this section, a description of the different indexes and parameters used in this work is presented.

The impact of the different degradation mechanisms which alter the performance of PV devices can be addressed through the Degradation Ratio (*DRatio*).

$$DRatio = \frac{J_{sc,degraded}}{J_{sc,reference}}, \tag{1}$$

where $J_{sc,degraded}$ is the short-circuit current density of a degraded PV device and $J_{sc,reference}$ is the short-circuit current density of a reference PV device of the same technology under identical conditions. Although this ratio neglects the influence of non-uniform degradation in a PV device, this approach has been considered appropriate for the purpose of this study.

In addition to the above, the short-circuit current of a reference PV device can be obtained through the following expression:

$$J_{sc,reference} = J_{sc,STC} \cdot \frac{G}{G_{STC}}, \tag{2}$$

where G is the actual global irradiance and the subscript STC refers to STC Conditions (irradiance: 1000 W/m², spectrum: AM 1.5G and cell temperature: 25ºC).

Taking this into account, it is possible to express the DRatio as:

$$DRatio = \frac{J_{sc,degraded}}{J_{sc,STC}} \cdot \frac{G_{STC}}{G}. \tag{3}$$

It should be noted that this equation could be considered similar to the one of the instantaneous Performance Ratio (PR$_{inst}$) index, expressed as:

$$PR_{inst} = \frac{P}{G} \cdot \frac{G_{STC}}{P_{STC}}. \tag{4}$$

According to the standard of the International Electrotechnical Commission, IEC 61724-1 [62], the PR is used to quantify energy losses due to a wide range of factors, such as irradiation, cell temperature, and degradation among others and it can be adapted to measure instantaneous losses by using power values. Furthermore, short-circuit current density (J$_{sc}$) can be used under the assumption that it decreases in the same extent as the maximum power. Therefore, if only the losses due to degradation are considered, neglecting the others mentioned before, it can be assumed that:

$$PR_{inst} = \frac{P_{degraded}}{G} \cdot \frac{G_{STC}}{P_{STC}} \approx \frac{J_{sc,degraded}}{G} \cdot \frac{G_{STC}}{J_{sc,STC}} = DRatio. \tag{5}$$

The solar spectrum and the spectral response of a PV device can be used to obtain the values of the current densities of equation (1):

$$J_{sc,reference} = \int_{\lambda_{min}}^{\lambda_{max}} E_G(\lambda) SR(\lambda) d\lambda. \tag{6}$$

$$J_{sc,degraded} = \int_{\lambda_{min}}^{\lambda_{max}} E_G(\lambda)\tau_{deg}(\lambda)SR_{deg}(\lambda)d\lambda, \tag{7}$$

where $E_G(\lambda)$ is the actual global spectrum (W m$^{-2}$ nm$^{-1}$), SR($\lambda$) and SR$_{deg}$($\lambda$) are, respectively, the spectral response of the non-degraded and degraded PV device (A/W), $\tau_{deg}(\lambda)$ quantifies the losses of the hemispherical spectral transmittance due to degradation, and $\lambda_{max}$ and $\lambda_{min}$ are, respectively, the longest and the shortest wavelengths (nm) of the device's spectral response.

The hemispherical transmittance is one of the most commonly used optical parameters to analyse the effect of the degradation on PV modules. It is defined as the ratio of the luminous flux transmitted by the surface of the PV module to the received luminous flux and it can be written as:

$$\tau = \frac{\Phi_e{}^t}{\Phi_e{}^i}, \tag{8}$$

where $\Phi_e{}^t$ is the luminous flux transmitted by the surface of the module (W) and $\Phi_e{}^i$ is the received luminous flux (W). The comparison of the values of transmittance before and after degradation provides an instant measurement of the spectral losses due to it (Equation 9):

$$\tau_{deg}(\lambda) = \frac{\tau_{degraded}(\lambda)}{\tau_{ref}(\lambda)}, \tag{9}$$

where $\tau_{degraded}(\lambda)$ and $\tau_{ref}(\lambda)$ are the spectral transmittance after and before degradation, respectively.

The spectral response (SR) of a PV module is the fraction of solar irradiance that is converted into current. SR is a function of wavelength ($\lambda$) and is related to the external quantum efficiency (EQE). It can be written as:

$$SR(\lambda) = EQE(\lambda) \cdot \lambda \cdot \frac{e}{h \cdot c}, \tag{10}$$

where SR ($\lambda$) [A /W] is the spectral response of the PV module, EQE ($\lambda$) is the external quantum efficiency, e is the electron charge [1.602176565 x 10$^{-19}$ C], h is Plank's constant [6.62606957 x 10$^{-34}$J s] and c is the speed of light [2.99792458 x 10$^8$ m/s].

The EQE is the relation between the number of electrons ($n_e$) accumulated at the external contacts of the module to the number of incident photons ($n_p$) [63]. It can be written as:

$$EQE(\lambda) = \frac{n_e}{n_p}. \tag{11}$$

As shown in Fig. 3, the spectral response varies with the PV technology [64–66]. This means that some technologies can be more or less sensitive than others to certain bands of the solar irradiance spectrum (Fig. 3).

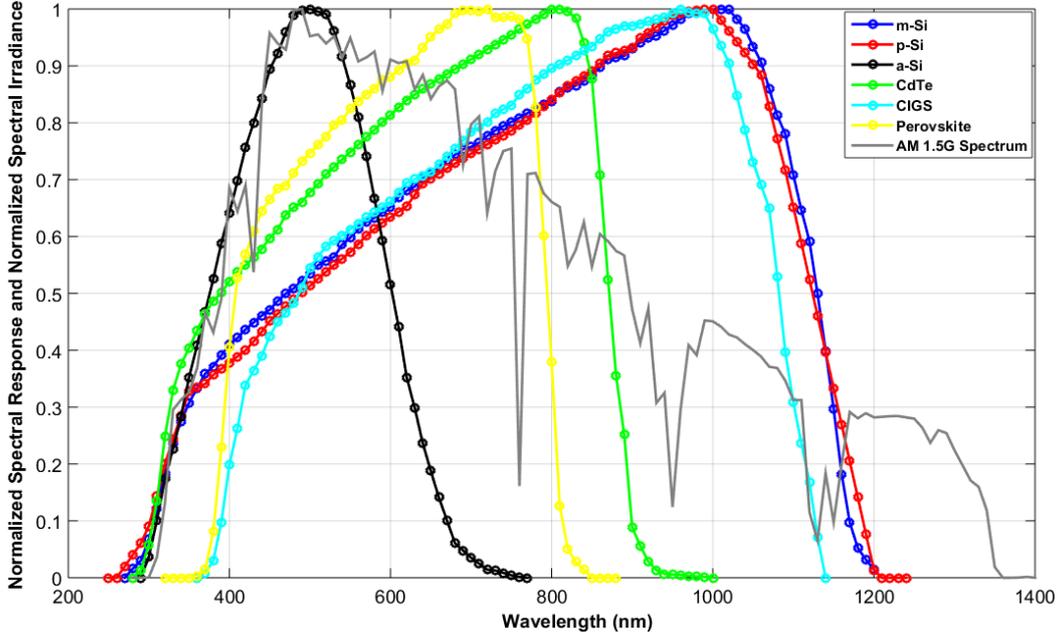

*Fig. 3. Normalized Spectral Response of different PV technologies and Normalized AM 1.5G reference spectrum. The data of the plotted curves were obtained from the EQE of commercial PV modules presented in: [67] (m-Si); [68] (a-Si); [69] (p-Si, CdTe and CIGS); and [4] (Perovskite)*

The DRatio, as defined in equations (1), (6) and (7), represents the global impact (broadband attenuation + spectral changes) of degradation on the irradiance profile. As a consequence, it takes into account both the different sources of degradation losses: (i) the light intensity reduction in the absorption band of the PV device (i.e. broadband attenuation) and ii) the effect of the non-horizontal spectral transmittance profile of the losses (i.e. the spectral changes).

If one wants to consider exclusively the broadband impact, equation (1) has to be rewritten as:

$$BDRatio = \frac{\int_{\lambda_{min}}^{\lambda_{max}} E_G(\lambda)\tau_{deg}(\lambda)d\lambda}{\int_{\lambda_{min}}^{\lambda_{max}} E_G(\lambda)d\lambda}, \tag{12}$$

where BDRatio is the Broadband Degradation Ratio, whose expression is obtained by the combination of equations (1), (6) and (7) and by considering a value of spectral response equal to 1 for every wavelength.

Besides, to consider only the spectral changes, equation (1) should be reordered in the following way:

$$SDRatio = \frac{DRatio}{BDRatio} = \frac{\int_{\lambda_{min}}^{\lambda_{max}} E_G(\lambda)\tau_{deg}(\lambda)SR(\lambda)d\lambda}{\int_{\lambda_{min}}^{\lambda_{max}} E_G(\lambda)SR(\lambda)d\lambda} \cdot \frac{\int_{\lambda_{min}}^{\lambda_{max}} E_G(\lambda)d\lambda}{\int_{\lambda_{min}}^{\lambda_{max}} E_G(\lambda)\tau_{deg}(\lambda)d\lambda}, \tag{13}$$

where SDRatio is the Spectral Degradation Ratio, which only takes into account the impact of the degradation losses spectral profiles on the performance of PV systems. Thus, the SDRatio has values above one in conditions of improved spectral performance (i.e. if the reduction in current is lower than that in irradiance).

Finally, it is important to highlight that the three indexes presented in this section can be used to estimate the power of a PV device as a function of the spectral and broadband degradation losses according to the following expression:

$$P_{degraded} = \frac{P_{STC}}{G_{STC}} \cdot G \cdot DRatio = \frac{P_{STC}}{G_{STC}} \cdot G \cdot BDRatio \cdot SDRatio. \qquad (14)$$

In this sense, the methodology described in this section along with the current literature allows to better understand the different degradation mechanisms. This makes it possible to improve the knowledge on the fundamental factors affecting the performance of PV systems. This is essential for energy yield assessments, and for addressing the degradation effects.

## 3.   Degradation mechanisms

The degradation plays a key role in energy performance of PV modules. Over the course of their lifetime, PV systems and modules can be affected by different adverse environmental conditions; [70], such as high temperatures, humidity or deposition and cementation of polluting particles. In Fig. 4, it can be seen which are the most habitual degradation mechanisms that affects PV modules worldwide. Among these, the effects of discoloration, delamination, corrosion and soiling on PV modules are addressed in this work. Furthermore, in this work, aging has been considered as an independent degradation mechanism itself and it has also been studied.  As a consequence, a depth search on current literature has been performed, with a special emphasis on the spectral effects. However, until now, soiling is the only mechanism whose spectral impact on PV modules has been widely documented, whereas few articles that address this issue for others degradation mechanisms, e.g., discoloration or aging, have been published. For this reason, in this review, a detailed table with some of the most relevant studies about the spectral effects of soiling on PV modules is presented.

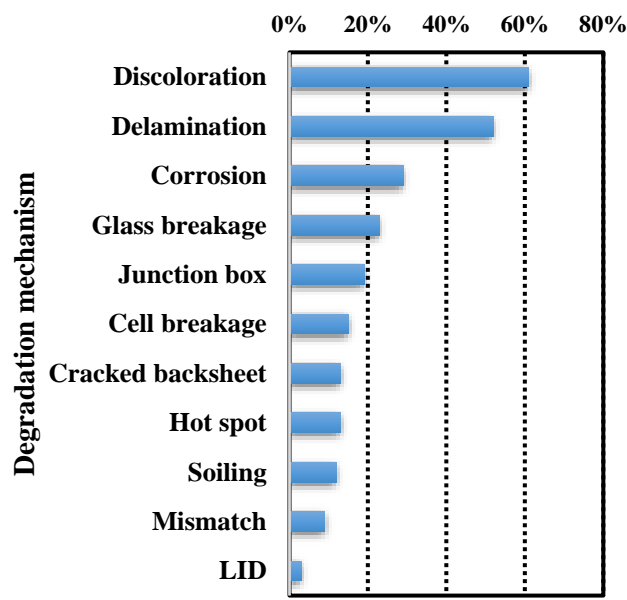

*Fig. 4. Observed changes in PV modules (percentage of PV modules that were affected with any degradation mechanism).*[Adapted from 71]

In next subsections, the degradation mechanisms commented before are described. First, they are generally assessed and then they are analysed from the perspective of their effects on PV spectral response.

### 3.1. Soiling

Soiling is defined as the accumulation of dust, particles, dirt and contaminants on the surface of PV modules. The soiling accumulated on the modules absorbs, reflects and deflects part of the sun radiation, reducing the intensity of the light reaching the PV cells and therefore the amount of energy converted by the module. Dust density accumulated on the surface of PV modules depends on a wide range of parameters, such as meteorological parameters (pollution, wind speed and direction, rainfall patterns…), location and tilt angle of PV modules, among others [72–75]. As commented, the main consequence of soiling is the reduction of the energy generated by PV modules [28]. The amount of energy losses due to soiling shows a strong correlation with the accumulated dust density [76–79] and with some properties of dust, such as chemical composition and size distribution [80]. In this sense, some studies show that PV efficiency is being affected more by finer particles in comparison with rough particles [81]. In the last decade, interest on the effect of dust on PV modules efficiency has increased exponentially. This has led to the performance of multiple studies and investigations around the world, and in particular in locations highly affected by soiling events, e.g., arid and desert regions of the Middle East, North Africa and the southwest of the USA [82–89]. In these locations, the effect of dust plays a fundamental role when evaluating the energy production and the revenues. One of the main focuses of interest is the ability to determine an optimal cleaning schedule [90–93] that makes it possible to maximize the economic profits. Another essential aspect is the choice of the appropriate cleaning method [94–98], considering in this case the physical and chemical properties of dust.

It is known that soiling produces both a reduction in the intensity of the sunlight and a change in its spectral profile. Qasem et al. [99] analysed the spectral response of different PV technologies with several soiling transmittance curves. The main conclusion of this study is that dust has spectrally selected effects on transmittance, being shorter wavelengths most affected. Therefore, the impact of dust on PV modules made with high energy-gap materials, such as CdTe and a-Si, is greater than the impact that it causes on c-Si and CIGS PV devices. This fact can be additionally influenced by the different layout of layers over solar cells in different technologies. Thus, while c-Si PV modules present a polymeric material layer under a glass cover, CdTe modules and certain types of a-Si modules present a glass-to-glass construction [60]. In [100], authors analysed the relationships that can be established between the reduction of efficiency and the variation of optical properties such as transmittance. In this work, the influence of dust on PV modules is addressed indirectly and the results of three different experiments are presented: in the first of them, the reduction in the transmittance of a glass coupon exposed outdoors, under roof or exposed to rain, with different tilt angles is used to analyse soiling losses; in the second one, the effectiveness of different coatings is studied by transmittance measurements with the same setup presented in experiment 1; in the last one, the influence of dust type over transmittance losses is analysed. The dust collected in experiment 1 was classified according its size and its chemical composition and was used to artificially soil different PV modules with the intention of comparing the reduction in transmittance of the coupons (Fig. 5) with the output power reduction of the modules. It was found that the difference between the transmittance and the power reductions due to soiling was limited to 2.5%. It should be noted that this difference is expected to change depending on the spectral transmittance of soiling, on the spectral distribution of the irradiance and on the spectral response of the cell [101]. In addition, the impact of dust on concentrator PV systems (CPV) through reflectance measurement has also been studied [102]. The main conclusion of this work is that soiling affects more CPV in comparison with flat PV modules. This is accounted for the additional losses in the optical elements that are part of this type of modules.

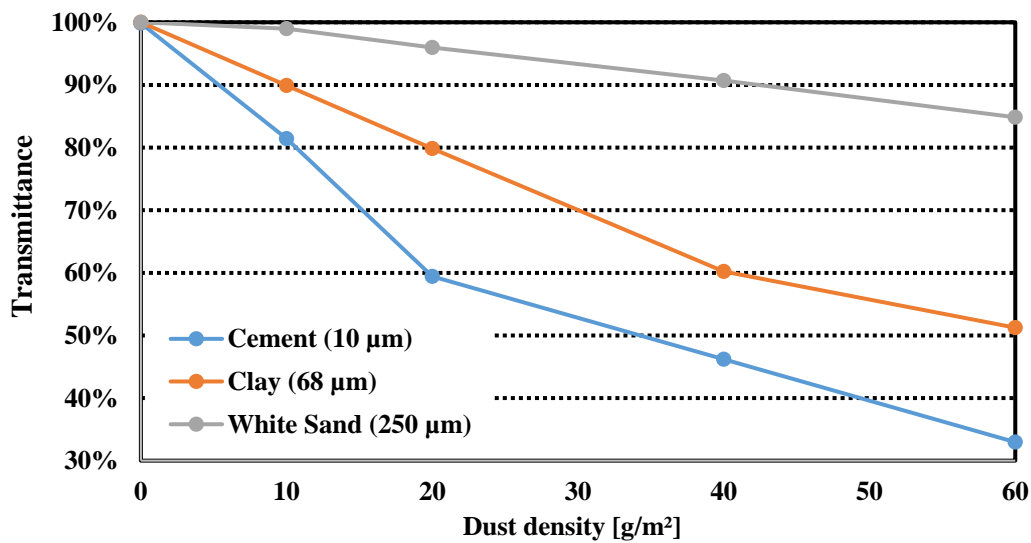

*Fig. 5. Transmittance for artificial soiling by different types of dust. Dust was collected using a set-up placed on the rooftop of the Department of Electrical Engineering at Catholic University of Leuven in Leuven, Belgium. [Adapted from 100].*

The study presented in [103] evaluates the spectral losses due to soiling for different PV technologies. For this purpose, the authors perform weekly transmittance measurements of a glass coupon that is exposed outdoors. The transmittance measurements are then used to estimate losses due to soiling that PV modules would experience if they were installed at the same location. The key finding of this study is the possibility to estimate soiling losses at a specific place by using transmittance measurements at a single wavelength. In addition, it was recently shown that the full soiling transmittance spectra could be modelled by a modified form of the Ångström turbidity equation [104]. In [105], reflectance and quantum efficiency (QE) measurements are used to determine spectral losses due to soiling in crystalline silicon PV modules. Reflectance measurements were performed outdoors using a portable spectroradiometer, whereas QE values were measured indoors. It was obtained that the average reflectance, considering wavelengths between 350 nm and 1100 nm, of moderately (3 g/m²) and heavily (74.6 g/m²) soiled PV modules increased by 58.4% and 87.2% respectively in comparison with the clean module. Moreover, the absorbance of the soil layer was calculated using reflectance and external QE measurements, and thus with reflectance and absorbance values, the spectral transmittance was determined solving the equation: transmittance + reflectance + absorbance = 1. Fig. 6 shows the transmittance, reflectance and absorbance of the dust layer. It can be appreciated a higher impact of dust on the transmittance at shorter wavelengths, as it was stated in [99] by Qasem et al. This issue is accounted for the authors of [105] by the occurrence of Mie scatter, that occurs mainly for small size particles where the transmittance attenuation of lower wavelength is more remarkable.

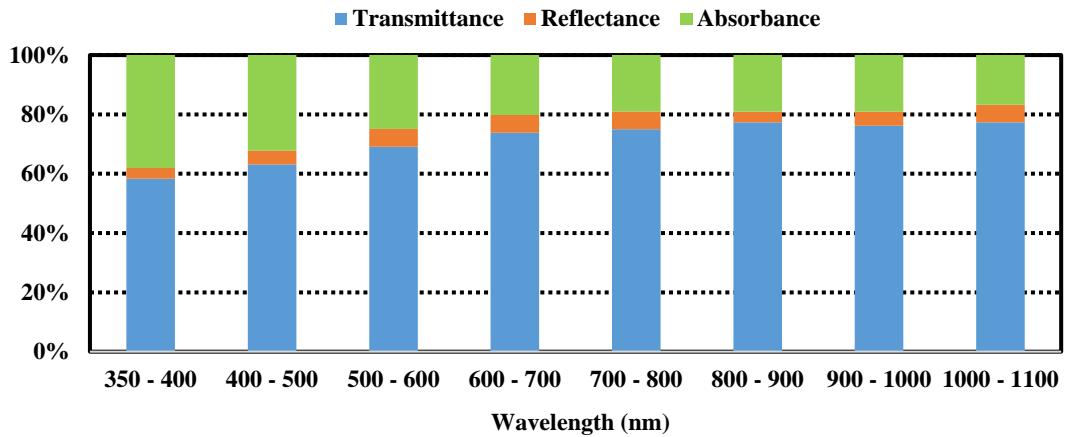

*Fig. 6. Reflectance, absorption, and transmittance spectra of the dust layer of the moderately soiled PV module. [Adapted from 105].*

These studies confirm that spectral measurements should not be neglected when analysing the impact of soiling on the PV performance. One of the purposes of this work is to summarize actual published studies about the effects of dust on the spectral response of PV modules. A review of the literature available about this topic is reported in Table 1.

*Table 1. Summary of dust effect on the optical properties of PV modules encapsulant and its associate impact on the modules performance.*

| Researchers [Reference number] | Location (Duration) | Spectral measurement | Key findings | Comments and Conditions |
|---|---|---|---|---|
| Mastekbayeva and Kumar [106] | Bangkok, Thailand (1 month) | Transmittance | Maximum dust accumulation of 3.7 g/m² during a one-month period during summer months.<br><br>12% reduction in optical transmittance. | Analysis of the impact of natural dust on the transmittance of a specific PV encapsulant |
| García et al. [107] | Tudela, Spain (16 months) | Irradiance | A model to calculate the irradiance on a tilted and soiled PV surface from the data of global and diffuse irradiance on a clean and horizontal surface.<br><br>On horizontal surfaces, soiling accounts for higher and more uniform energy losses during the different seasons of a year. | Analysis of optical losses due to soiling and radiation angle of incidence (AOI).<br><br>Setup: 3 calibrated PV cells are installed aligned with the modules and another 3 cells are mounted horizontally. In this way, AOI losses were registered.<br>The irradiance measurements that |
| | | | For tracking systems, daily energy losses are up to 8%. For horizontal surfaces, expected losses up to 22% can be observed. | have been taken with the PV cells are compared with the irradiance measured by a pyranometer. |
| Qasem et al. [99] | Kuwait (1 month) | Transmittance | Dust mainly affects lower wavelengths, so the effect on PV spectral response depends on the PV technology.<br><br>Higher energy-gap technologies, such as a-Si and CdTe are more affected than c-Si technology. | Spectral transmittance and dust density is correlated.<br>The impact of dust on different PV technologies is analysed through the effective spectral response. |
| John et al. [108] | Mesa, USA (Not available) | Reflectance and quantum efficiency (EQE) | The heavily soiled cell presents high reflectance as compared to the medium and lightly soiled at all wavelengths before any cleaning technique is applied. The reflectance is maximum at 1000 − 1100 nm wavelength range with values around 50 % for the heavily soiled cell.<br><br>The reflectance spectra decreased after each cleaning step.<br><br>The heavily soiled solar cell has negligible QE between 350-600 nm wavelengths.<br><br>After the first 2 steps of cleaning with compressed air, the QE curve increases significantly, however after the water clean, the QE curve decreases by 1 %. | Setup: 3 cells of a PV module were artificially soiled with different levels of soil density (74,6 g/m² (heavy soiled), 3.18 g/m² (medium soiled) and 0.6 g/m² (lightly soiled)).<br><br>Measurements of reflectance and quantum efficiency were performed before and after different clean techniques, divided in three steps. |
| Yilbas et al. [109] | Saudi Arabia (Not available) | Transmittance | The formation of mud on the glass samples influences | Setup consists on PV protective glass samples. |

| | | | their properties, such as, absorption, transmittance and microhardness. | Analysis of the influence of dust characteristics and mud formation on the transmittance of the glass samples. |
|---|---|---|---|---|
| | | | A relevant spectral transmittance reduction is noticed after the removal of the mud. The main causes of this are linked with: (1) the rest of mud that cannot be eliminated from the glass surface and (ii) chemical reactions that cause changes in the properties of the glass. | Different situations of dust and mud accumulation and formation were artificially reproduced in a laboratory (dust thickness was measured during a sand storm in Saudi Arabia over 2 weeks) |
| John et al. [105] | Mesa, USA (5 years) | Reflectance and external quantum efficiency (QE) | The characterization of spectral and angular losses of PV modules was performed by using an innovative method, which was tested with different values of dust density. | QE measurements were performed indoor and reflectance measurements were done outdoors. |
| | | | Reduction of 41% in short circuit current at dust density of 74 g/m². | Setup: 1 p-Si module was installed horizontally outdoors during 4 years, showing different levels of soil density, to perform spectral studies. |
| | | | The average absorption is 61% at dust density of 74 g/m². | 3 soiled and 1 cleaned p-Si modules were installed for AOI studies. |
| | | | 7% of the light is transmitted through a dust layer of 74 g/m². | An equation was used to estimate the absorption spectra in wavelengths between 350 nm and 1100 nm. |
| | | | EQE reduction is strongly dominated by the dust absorption. | The spectral transmittance was obtained from the reflectance and the absorption values. |
| Burton et al. [110] | Simulations | Transmittance | A method to model the spectral | The spectral effects of different types of |
| | | | transmittance through various types of dust, which correspond in colour to naturally occurring soils in US southwest, on triple-junction solar cell | soil, which were generated using mineral pigments and traces of soot, on a calibrated isotype cell were calculated. |
| | | | | Dust was considered as a stand-alone optical element of a CPV system. |
| | | | | The limitation of the current of a high concentrator PV device was calculated through the changes in its spectral response. |
| Boyle et al. [111] | Colorado, USA (5 weeks) | Transmittance | 4.1% decrease in transmittance per 1 g/m² of dust accumulated on the glass sample surface. | Data collection from 2 different locations in Colorado. |
| | | | The highest accuracy of this correlation has been found for angles of incidence between 20º and 60º, and for dust density values up to 2 g/m². | Identical setup at each location. Glass plates installed under a roof with different orientation and tilt angles. |
| | | | | Transmittance measurements were performed on clear sky days. Only measurements from 375 nm to 1150 nm have been considered. |
| Tanesab et al. [112] | Perth, Australia (Not available) | Transmittance | A dust density of 0.04 mg/cm² decreases transmittance by a factor of 18.34%. | Measurements of the transmittance of a glass coupon with different levels of dust density. The dust was collected at the field location of the PV plant. |
| | | | | Chemical and physical properties of dust were also investigated. |
| Pedersen et al. [113] | Norway (2 months) | Transmittance | Transmittance is reduced by 0.09% and 0.11% per 10 | A linear correlation between transmittance and |

|  |  |  | mg/cm² for the normal glass coupons and anti-soiling (AS) glass coupons, respectively. Coupons with AS coating accumulates more soil than the normal coupons. | dust density is presented. Glass coupons are exposed with the same tilt angle of PV modules to analyse dust density. Some of them had coatings applied. |
|---|---|---|---|---|
| Abderrezek and Fathi [85] | Tipaza, Algeria (Not available) | Transmittance and solar spectrum | Indoor tests: A fast decrease of spectral transmittance is due to the presence of cement and ashes in the environment. Finer particles affect more transmittance than coarser particles.<br><br>Outdoor tests: Reflectance, absorption and light scattering caused by dust are the main factors that contribute to the degradation of the solar spectrum.<br><br>Results obtained from indoor and outdoor tests are quite similar. | The impact of dust on the transmittance of PV encapsulant was analysed by the performance of indoor tests and the use of several types of dust.<br><br>Outdoor tests: Transmittance measurements of glass coupons exposed outdoors were performed to evaluate the effect of dust on the solar spectrum that reaches PV cells. |
| Micheli et al. [114] | 8 different locations (8 weeks) | Transmittance | Average weekly losses as high as 3.2% in hemispherical transmittance and 11.9% in direct transmittance were registered in Jaén, Spain and in Tezpur, India, respectively.<br><br>Maximum losses during the study period of 7.8% in hemispherical transmittance and 46.9% in direct transmittance were measured in Tezpur, India and in El Shorouk City, Egypt, respectively.<br><br>Transmittance losses are higher in the UV | Identical glass coupons were exposed horizontal outdoors at different locations worldwide to analyse the spectral impact of soiling on PV modules. |

| | | | | region of the spectrum. A linear relation between the area covered by dust particles and hemispherical transmittance was discovered |
|---|---|---|---|---|
| Paudyal et al. [115] | Kathmandu valley, Nepal (5 months) | Transmittance | A regressed model that relates meteorological variables to dust density. An inverse linear dependence between dust density and transmittance losses, and a linear correlation between transmittance losses and power reduction. 29.76% decreases in the power output. | p-Si technology. Transmittance measurements at 3 different wavelengths: 450 nm, 600 nm and 750 nm. |
| Ravi et al. [116] | Indoor experiments | Reflectance | A method to evaluate the effectiveness of anti − soiling (AS) coatings. Transmittance values are used as indicators to compare the different coatings. | Setup: A soil deposition chamber is used to evaluate different AS coatings. A 9-cell m-Si module and 3 one-cell modules were used. Dust was collected from a PV plant in Mesa, USA. Reflectance measurements were performed and transmittance values were approximated by short circuit current measurements. |
| Goosens [117] | Indoor experiments | Transmittance | A method to evaluate the benefits of anti − soiling (AS) and anti-reflective (AR) coatings in order to mitigate the impact of dust accumulation above PV modules | Setup: A wind tunnel was used to simulate natural deposition processes of soil on PV modules surface. The transmittance |
| | | | surface. | of glass coupons placed horizontally was measured to compare the effectiveness of the coating under different conditions of: wind speed, humidity, dust density and dust properties. |
| Bengoechea et al. [118] | Indoor experiments | Transmittance | Power losses in CPV systems were higher than losses obtained for m-Si technology, for the same soiling conditions (i.e. for a dust density of 11.16 g/m² of salt, power losses for m-Si technology reach a value of 1.4%, while for CPV they increased up to 7.1%). | Comparison between the values of direct and hemispherical transmittance of different types of dust and the reductions in power for CPV and monocrystalline silicon technologies. Procedure for power losses estimation: 1) Short-circuit current density was addressed by using the spectral response of each technology, and the transmittance of dust. 2) Power was calculated assuming that it was proportional to the short-circuit current density (in the case of CPV technology, multijunction solar cell, power loss reduction was considered to be the same as the decrease of the lowest short-circuit density. |
| Ilse et al. [119] | Doha, Qatar (28 days) | Transmittance | Long periods of outdoor exposure cause important drops in the spectral transmittance of glass coupons. A dependency of wavelength is | Several experiments were conducted under outdoors conditions to discover what were the mechanisms of dust layer formation on PV glazing. |
| | | | observed: transmittance losses are higher for shorter wavelengths (< 600 nm). | Glass samples were exposed outdoors during different time periods so as to collect different levels of soiling. |
| | | | | Transmittance values of soiled glass samples were correlated with PV energy losses. |
| Fernández et al. [61] | Jaén, Spain (1 year) | Transmittance | First investigation related with the impact of soiling on the spectral response of multijunction-based CPV systems under outdoor conditions. | Setup: A glass coupon was installed horizontally outdoors to collect natural dust, while a clean coupon was store to be used as the baseline. Typical meteorological parameters and the spectrum of the direct normal irradiance were measured with an atmospheric station and a solar spectral irradiance meter respectively. |
| | | | Some indexes are defined to evaluate specifically the impact of the soiling transmittance profile on the short-circuit current density of CPV modules, such as the Spectral Soiling Ratio (SSR) and the Soiling Mismatch Ratio (SMR) that considers the soiling spectral effects on the current balance among both, the top and the middle subcell. | Procedure: Weekly direct transmittance measurements of both coupons were performed in order to obtain the value of soiling transmittance. The values of the indexes (SSR and SMR) are calculated using different equations with the recorded data and taking into account the response of a triple-junction PV cell. |
| | | | Soiling increases annual transmittance losses by an average of 2% | |
| | | | Transmittance losses in the top subcell due to soiling are 4% higher in comparison with these obtained in the middle. | |
| Tanesab et al. [120] | Indoor experiments | Transmittance | For the same density value, dust from Babuin and Perth did not have a significant difference in their impact on the performance of each PV technology. | Dust collected at 2 different locations (Perth, Australia and Babuin, Indonesia) was used to artificially soil PV modules of different technologies; i.e. m- |

| | | | | Si, p-Si and a-Si to analyse the effects of dust with different properties (morphology and chemical composition) on their performance. |
|---|---|---|---|---|
| | | | PV modules of different technologies showed similar power losses affected by each type of dust. This can be justified by the flat spectral transmittance curves of both types of dust. | |
| Micheli et al. [101] | Jaén, Spain (1 year) | Transmittance | Two or three single wavelengths measurements can be used to model the full soiling transmittance spectra. For each PV material is possible to identify an optimal combination of wavelengths that minimize the modelling error. | Soiling detection was analysed by using hemispherical transmittance measurements of a glass coupon exposed outdoors. |
| Smestad et al. [104] | 8 different locations (8 weeks) | Transmittance | The soiling transmittance spectra could be modelled through a modified form of the Angstrom turbidity equation [121]. The particle size distribution could be described through the IEST-STD-CC 1246E cleanliness standard. A linear correlation between area coverage and transmittance was found. | Identical glass coupons were exposed horizontal outdoors at different locations worldwide to analyse the spectral impact of soiling on PV modules. |

### 3.2. Discoloration and Delamination

Discoloration and delamination (D&D) of encapsulant in PV modules are two degradation mechanisms that reduce the electrical performance because they diminish the transmittance of the encapsulant of the PV module, which it often is Ethyl vinyl acetate (EVA), by modifying its chemical and physical properties due to outdoor exposure. This attenuation produces a drop of the short-circuit current ($I_{sc}$) of the PV module and thus a reduction of its efficiency. This two degradation mechanisms are the most observed failures in PV modules in field systems [71,122,123] as it was previously shown in Fig. 4. The data plotted in this figure were collected from four different PV plants located in a hot-dry desert climate region [122]. They show the large number of PV modules that manifest signs of discoloration and delamination, since 61% and 52% of the total installed modules were affected by these two mechanisms respectively.

Discoloration, also known as browning, can be visually detected because of the change in colour of the PV encapsulant material. This occurs as a consequence of the photo-thermal degradation of EVA encapsulant under high temperatures and high values of UV radiation (Fig. 7). The magnitude of the degradation follows a relationship with the colour of the EVA (the darker the EVA colour, the greater the degradation). The final effect of discoloration is the reduction in PV module efficiency due to the decrease in light transmittance. Fig. 8 shows how discoloration affects the transmittance causing performance losses. The measurements were performed on a 25-year-old monocrystalline-silicon (m-Si) module exposed in the coastal area of Yeosu, South Korea. A secondary effect of discoloration is the production of acetic acid [124], which get locked inside the module at different layers. This causes the accumulation of gases that, in turn, can lead to the formation of bubbles or the EVA delamination. Another effect of acetic acid is the corrosion of metallic contacts, which can cause an increase of the shunt resistance value. It should be noted that discoloration only affects PV modules with a polymeric PV top-sheet encapsulant, which is typical in m-Si and p-Si technologies.

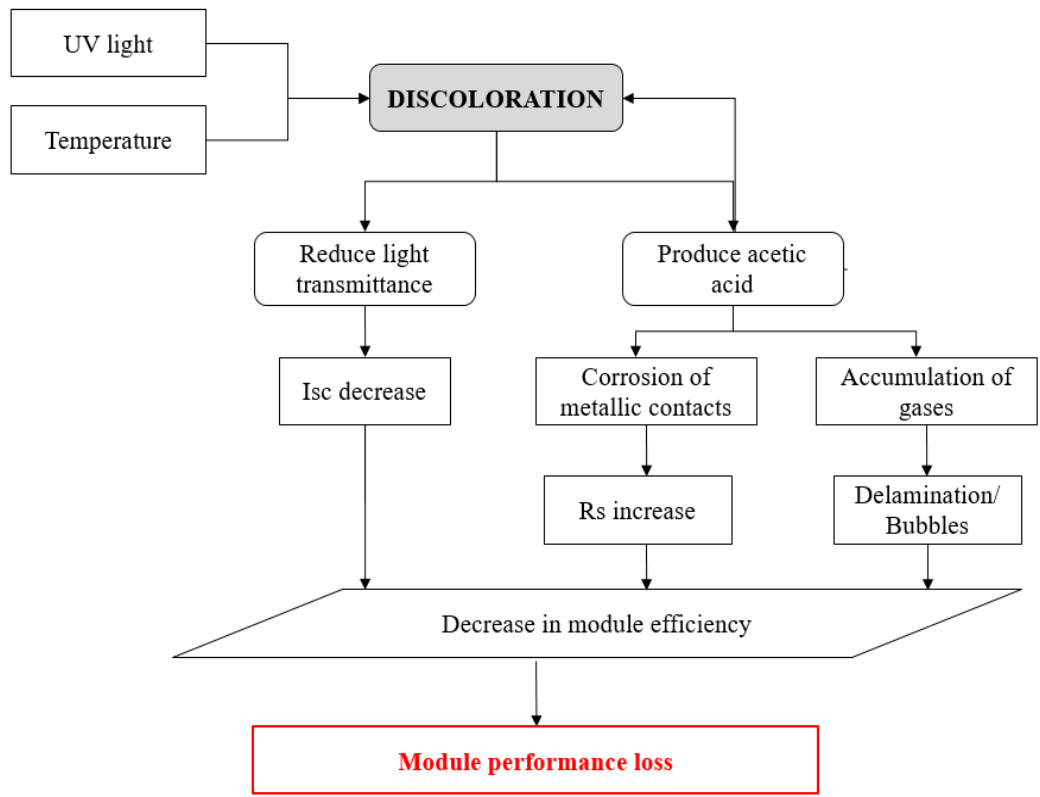

*Fig. 7. Main causes and effects of discoloration in PV modules.* [Adapted from 124].

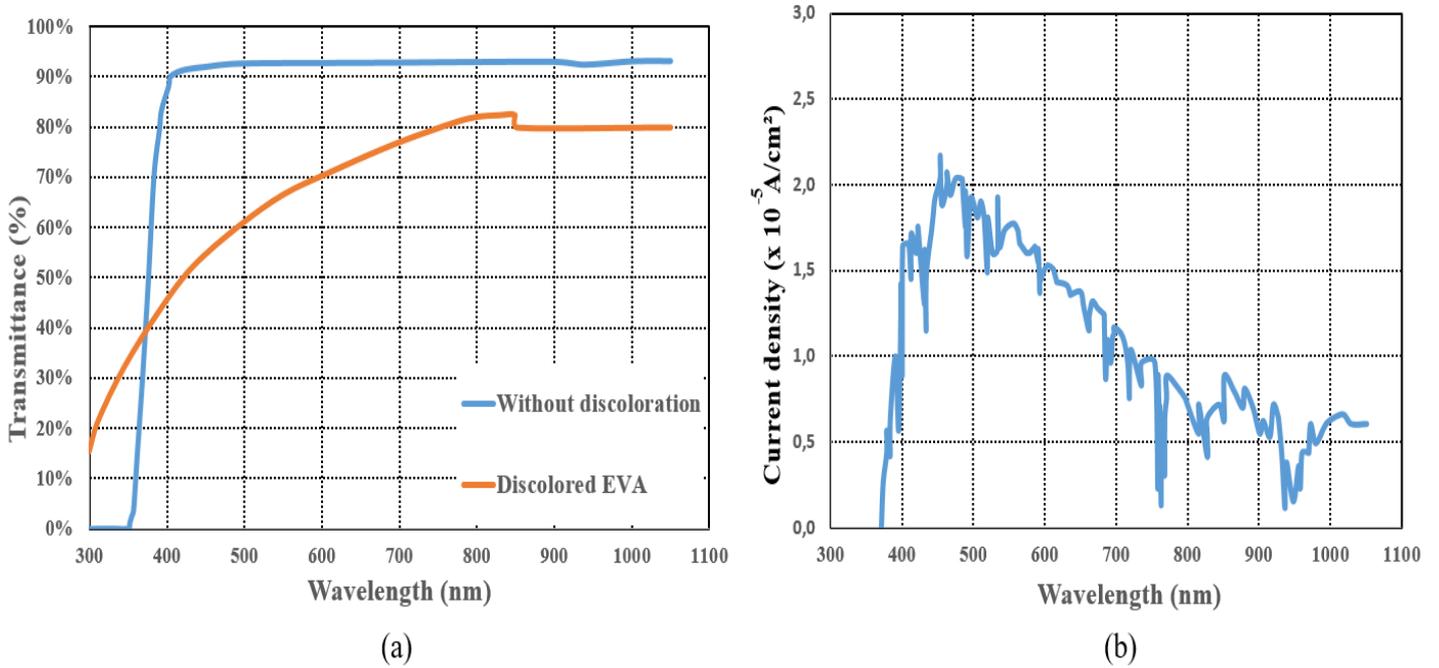

*Fig. 8. Discoloration effect on transmittance; (a) transmittance of EVA with and without discoloration, (b) transmittance induced current loss* [Adapted from 125].

In [38], reflectance measurements are performed to analysed discoloration. A parameter called Yellowness Index (YI) is used to evaluate the magnitude of the degradation. This parameter could be correlated with power losses as it is shown in Fig. 9.

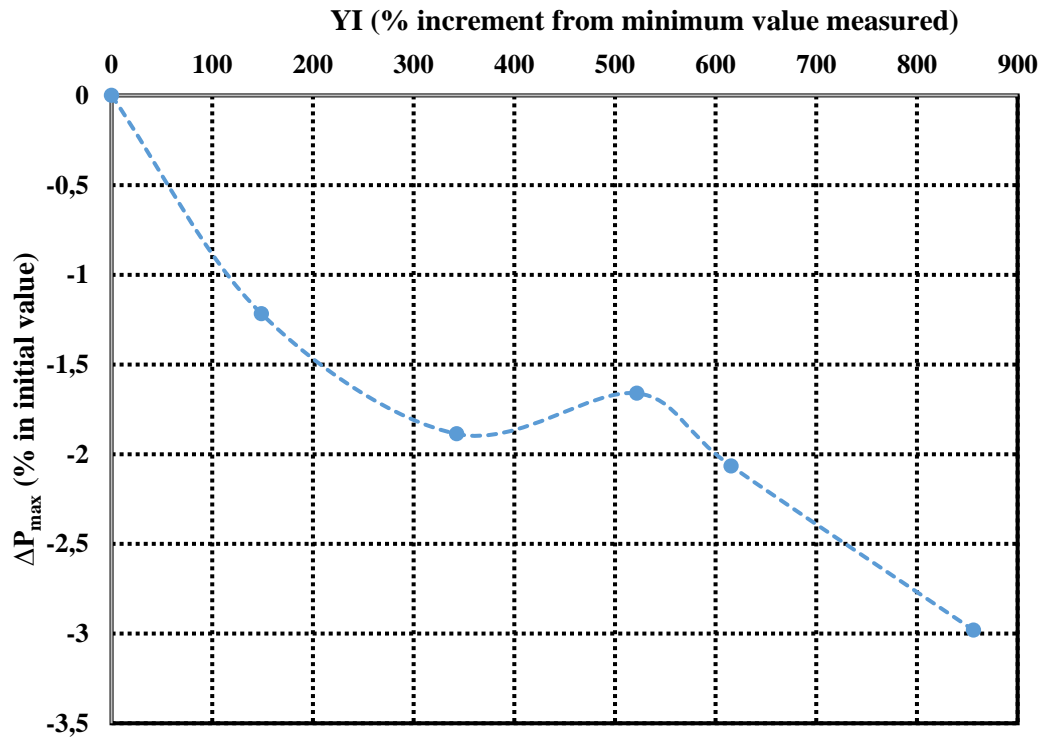

*Fig. 9. Change in the maximum power (in % of the initial value) of a PV solar module vs YI. The module was subjected to an Ultraviolet test in a UV chamber, reaching a maximum of 195 kWh/m² of UV light.[Adapted from 126].*

Delamination is another degradation mechanism related with module encapsulant (EVA). As commented above, the delamination of the EVA usually appears simultaneously with discoloration due to the formation of acetic acid that gets trapped within the different interfaces of the PV modules enhancing the probabilities of bubbles formation and delamination [124]. It occurs as a result of a weakening chemical adhesion between the EVA and the active part of the module or between the EVA and the glass. Like discoloration, it reduces the transmittance and increases the reflectance, because of a phenomenon known as light decoupling [125], which causes the separation of light beams when they reach the surface of PV modules. Another effect is the acceleration of corrosion, due to the fact that moisture can penetrate into the cell encapsulation more easily. Fig. 10 shows a m-Si module with delamination.

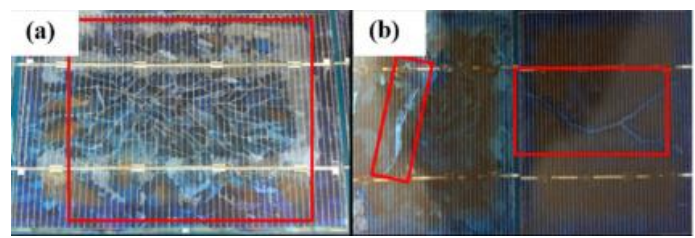

*Fig. 10. Delamination of EVA and cracks in solar cells of a m-Si PV module after 22 years of outdoor exposure in a location in China with a tropical climate. [127]*

As it has been mentioned previously, delamination is strongly linked to corrosion [125], which is discussed in the following subsection.

### 3.3. Corrosion

Like D&D, corrosion is another mechanism associated with the degradation of the PV modules encapsulant and it only affects PV modules that have a polymeric one, which is typical of m-Si and p-Si technologies. It could occur between different PV cells or between the cells located on the sides of the PV module and the frame. The main causes of corrosion [48] are: (1) the ingress of moisture through module edges or as a consequence of the existence of delamination (Fig. 11), (2) the presence of acetic acid and (3) the weak adhesive bonds at the interfaces (cell/encapsulant, glass/encapsulant, backsheet/encapsulant). In addition, the risk of corrosion increases in locations with high temperatures and humidity.

Two different types of corrosion can be distinguished: (1) chemical corrosion, which occurs at night, by chemical reactions in the metallic and semiconducting elements; and (2) electrochemical corrosion, which occurs during the day, whose magnitude depends on the sunlight radiation, which can lead to high voltage values.

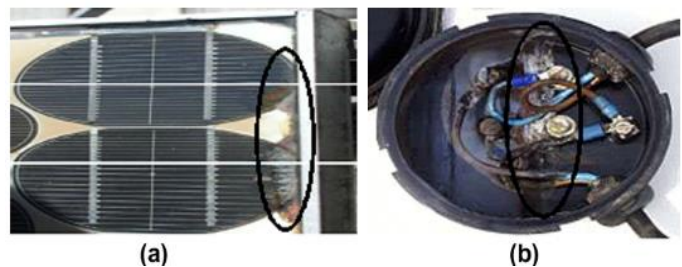

*Fig. 11. Visual consequences of corrosion in a PV module: (a) edge and (b) junction box [128]*

The main effects of corrosion are the appearance of leakage currents, as a consequence of a raise in the value of series resistance of the PV module and the formation of hot spots. Effects over the spectral response have not yet been studied; however, as corrosion is promoted by D&D mechanisms [125,128,129], the spectral effects caused by them will be probably increased if corrosion is also occurring.

Fig. 12 shows the most common degradation mechanisms related with encapsulant of silicon PV modules which were installed in three different PV plants located in places with diverse climates: Arizona, US (desert climate), Seoul, South Korea (humid continental climate) and Miami, US (tropical climate) [130]. Furthermore, some studies [128,131,132] claimed that D&D and corrosion are among the most common degradation mechanism for the whole PV module.

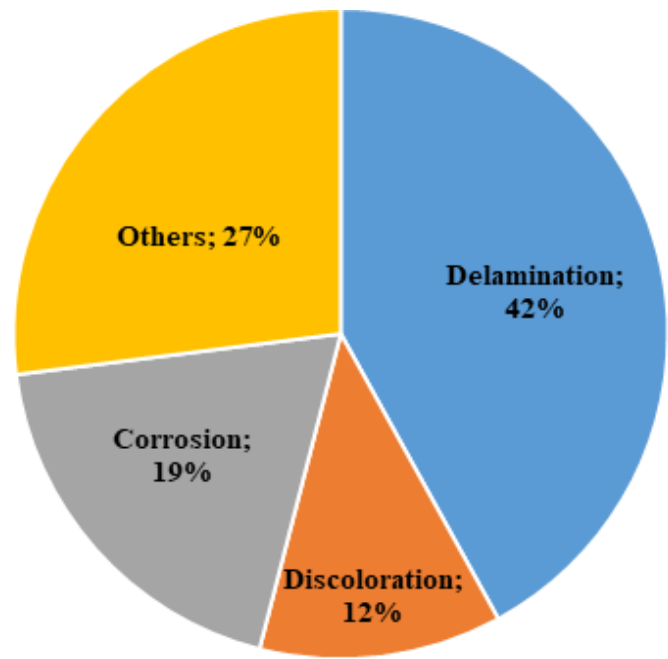

*Fig. 12. Silicon PV modules: representativeness of degradation mechanisms related with the encapsulant.* [Adapted from 128].

### 3.4. Aging

Outdoor exposure over time of PV modules causes the progressive degradation of their efficiency. Although, some studies consider that aging is not a degradation mechanism [27,133]; in this work it has been treated as an independent one. This choice can be accounted for the results presented in [134], where the existence of others signs of degradation, such as the discoloration of the top-sheet encapsulant, did not reflected in the same way across all the 20 year old silicon modules that comprised the PV system. In this way, aging can be analysed by taking into account only the effects of weathering over time. Furthermore, it is widely reported that the exposure of PV modules to adverse environmental conditions (extreme temperatures, heavy rains, higher humidity values and the presence of contaminants) promotes the appearance of other degradation mechanisms analysed previously, such as D&D [135,136] and corrosion [27,137]. In contrast with these three mechanisms, aging is not expected to be related with the change of the optical properties of the module encapsulant, rather to the inherent degradation of the spectral response of the solar cells that comprise a module.

Numerous studies [133,134,137–139] analysing the reduction of efficiency of PV modules due to aging have been published. In the study published by Ishii et al. [138] the value of the instantaneous Performance Ratio (PR) was calculated during 4 consecutive years for different PV technologies in Tsukuba, Japan (Fig. 13). Some conclusions of this study are: (1) m-Si modules degrade in a higher rate than p-Si and a-Si:H/c-Si modules; and (2) the PR of the CIGS modules improves during the first year of operation because of light soaking.

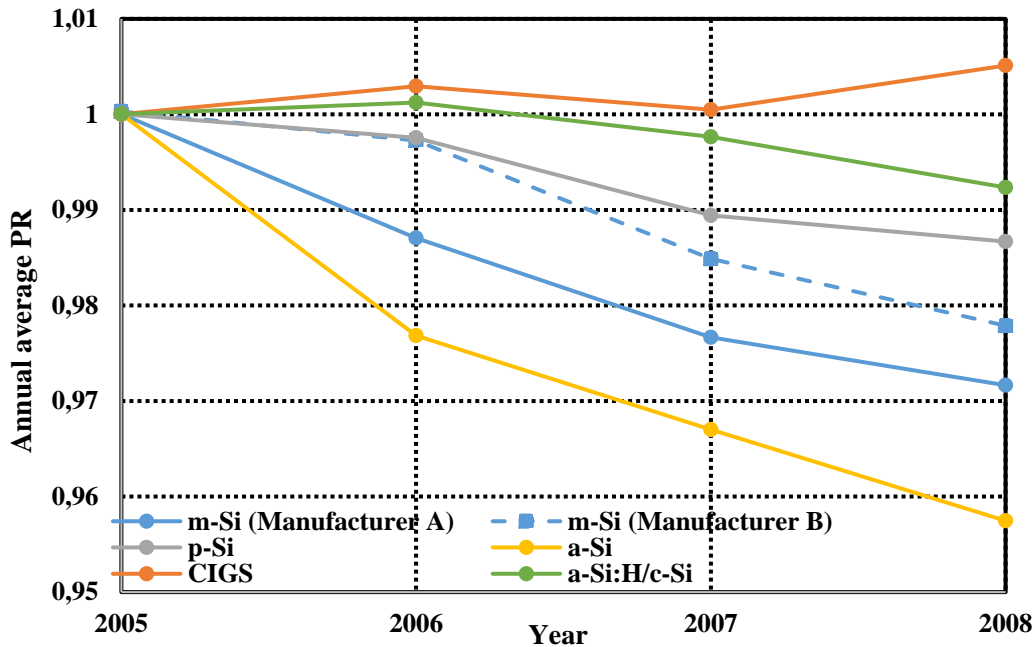

*Fig. 13. Annual average performance ratio corrected to 25ºC from 2005 to 2008 for different PV technologies installed in Tsukuba, Japan* [Adapted from 138].

The study performed by Parretta et al. [135] shows the effects of aging on reflectance measurements for four different m-Si and p-Si PV modules with and without anti-reflective coating (ARC) from various manufacturers. The main outcomes are: (1) A wider dispersion of reflectance measurements was detected in aged PV modules, which makes the incoming light distribution optically more homogeneous; (2) a clear correlation between electrical and optical degradation exists, but its magnitude strongly depends on the manufacturer that produces the modules, and as a consequence of this; (3) the four PV modules did not degrade consistently. For these last two reasons, the results presented in this study should not be considered universally valid for all the m-Si and p-Si PV modules; nevertheless, the methods developed by the authors to evaluate the optical degradation of aged modules can be considered as a guideline for the PV community.

Han et al. [127] analysed the degradation of m-Si PV modules with an EVA encapsulant after 28 years of outdoor exposure in a tropical climate. After a visual inspection, they found that 100% of the modules showed signs of EVA discoloration and bus bar corrosion. The degradation of the electrical performance was conducted by measuring the IV curves of the modules under STC conditions using a solar simulator before and after replacing the junction boxes of modules with 22 years of outdoor exposure. The main outcome of these measurements was that, after 28 years of operation, the short-circuit current ($I_{SC}$) presented the highest degradation rate, with an average of 14.3% compared to the nameplate value. On the other hand, the degradation rates of the maximum power and open-circuit voltage were 6.5% and 1.5%, respectively. The authors concluded that the primary cause of power loss was the drop in the $I_{SC}$ due to the degradation of the optical properties of the EVA encapsulant. EVA aging was mostly found in PV modules with its discoloration or yellowing. Its impact was evaluated by measuring the spectral transmittance in the center and at the edge of the cells of the module (Fig. 14).

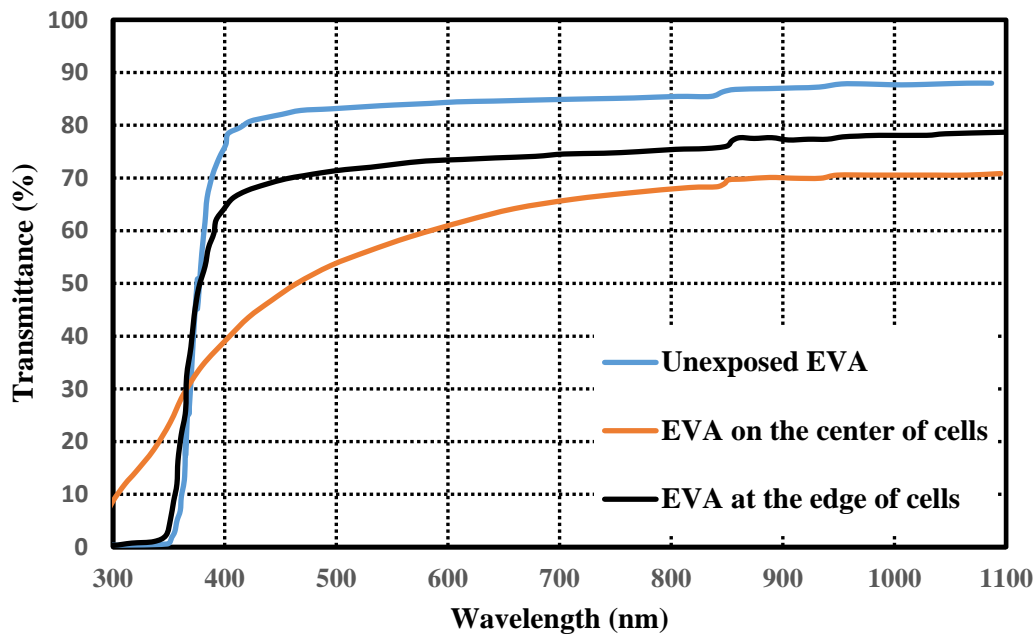

*Fig. 14. Transmittance of EVA encapsulant in different parts of the PV modules after 30 years of exposure to a tropical climate* [Adapted from 127].

In both cases, on the center of the cells and at the edge, the transmittance is smaller for wavelengths between 380 nm and 1100 nm in comparison with the transmittance of the unexposed EVA. In particular, the transmittance measured on the center presented the highest decrease in this range. On the other hand, between 300 nm and 380 nm, the transmittance on the center was even higher than the transmittance of the unexposed EVA. This was justified by assuming a total consumption of the UV absorbers in this region. Using the data plotted in Fig. 14 and the AM1.5 solar spectrum, the short-circuit density loss was calculated by the authors of that study, obtaining losses of 10.1% and 15.1% at the edge and on the center, respectively. These values were in line with those obtained after evaluating the degradation of the electrical performance. With these results, the authors of the study concluded that the main source of power losses was the degradation of the optical properties of the EVA encapsulant due to aging.

In [140], the effect of aging in different encapsulating materials for PV modules was analysed through damp heat tests, based on the exposure of these materials to high humidity and temperature values. The evaluation of the aging effect was done through infrared spectroscopy in attenuated total reflectance (ATR) mode of the infrared absorption spectra and through other thermal and mechanical measurements.

As it has been mentioned before, there are only three studies [127,135,140] that address aspects related with the impact of aging on the transmittance changes of PV encapsulant. For this reason, further investigations should be performed in order to acquire relevant results which help manufacturers to improve the design of PV encapsulant films and coatings.

## 4.   Spectral impact analysis

A detailed analysis of the impact of two of the degradation mechanisms mentioned before (soiling and discoloration) on the spectral response of PV modules of six different technologies is presented in this section. The lack of spectral data on the remaining aforementioned degradation mechanisms does not make an in-depth study possible.

This analysis is based on the estimation of the spectral impact of degradation mechanisms by taking into account multiple spectra with different values of air mass (AM), aerosol optical depth (AOD) and precipitable water (PW). It has been demonstrated that these three atmospheric parameters are the most influential on the spectral irradiance in clear-sky days, and they are widely used to analyse the effect of solar spectrum variations on the performance of PV devices [141–144]. Table 2 includes the wavelength limits of the absorption band of each material along with the limits of the different spectral regions that have been considered in this study.

*Table 2. Wavebands considered in the analysis.*

|  | **Waveband** | **$\lambda_1$ [nm]** | **$\lambda_2$ [nm]** |
|---|---|---|---|
| Spectral regions | Ultraviolet (UV) | 280 | 400 |
|  | Visible (VIS) | 400 | 700 |
|  | Near-infrared (NIR) | 700 | 1240 |
| PV material absorption bands | Monocrystalline silicon (m-Si) | 280 | 1200 |
|  | Polycrystalline silicon (p-Si) | 280 | 1200 |
|  | Amorphous silicon (a-Si) | 290 | 770 |
|  | Cadmium telluride (CdTe) | 290 | 1000 |
|  | Copper indium gallium diselenide (CIGS) | 360 | 1140 |
|  | Perovskite | 360 | 840 |

AM can be defined as the relation between the optical path length through the atmosphere and the optical path length at zenith. Its value only depends on the position of the sun, with a minimum value of 1 (zenith angle = 0º) and a maximum value of 38 (zenith angle = 90º) [145]. The changes of AM coefficient have a relevant impact in the solar spectrum due to scattering and absorption. The influence of AM in the global spectral irradiance is shown in Fig. 15-top. The ultraviolet (UV) region of the spectrum is the region that experiences the larger attenuation with a rise of AM.

The Aerosol Optical Depth (AOD) parameter is used to quantify the impact of aerosols on the irradiance. Aerosols are solid or liquid particles suspended in the air that scatter and absorb sunlight. Its impact on the spectral irradiance can be estimated by the Ångström turbidity formula [146] as AOD = $AOD_{500}$ $(\lambda/0.5)^{-\alpha}$. In this formula, $AOD_{500}$ accounts for the quantity of aerosols in a column of the atmosphere at 500 nm, $\lambda$ and $\alpha$ are the wavelength and the Ångström exponent respectively. The influence of AOD at 500 nm in the incoming global spectral irradiance is represented in Fig. 15-middle. As it can be seen, it mainly affects UV and visible (VIS) regions of the spectrum. However, the extent of its impact is shorter in comparison with AM impact.

The precipitable water (PW) parameter is used to evaluate the amount of water vapour contained in the atmosphere. It quantifies the liquid water that would be obtained if all the water vapour in the air was condensed. The influence of PW is plotted in Fig. 15-bottom. PW mainly affects the near-infrared (NIR) region of the spectrum.

Considering the effects mentioned above, it can be stated that the changes of each parameter will cause a different impact on the performance of PV devices dependent on their absorption bands. In the following two subsections, the spectral impact of soiling and discoloration on different PV technologies is addressed through the consideration of real transmittance degradation data and assuming that the spectral response of the PV device does not vary (equation 13). It should be noted that the analysis presented in this section only considers how changes in the transmittance of the PV encapsulant impacts the spectral behaviour of PV devices, as no data concerning the possible changes of the spectral response of solar cells was available at this stage.

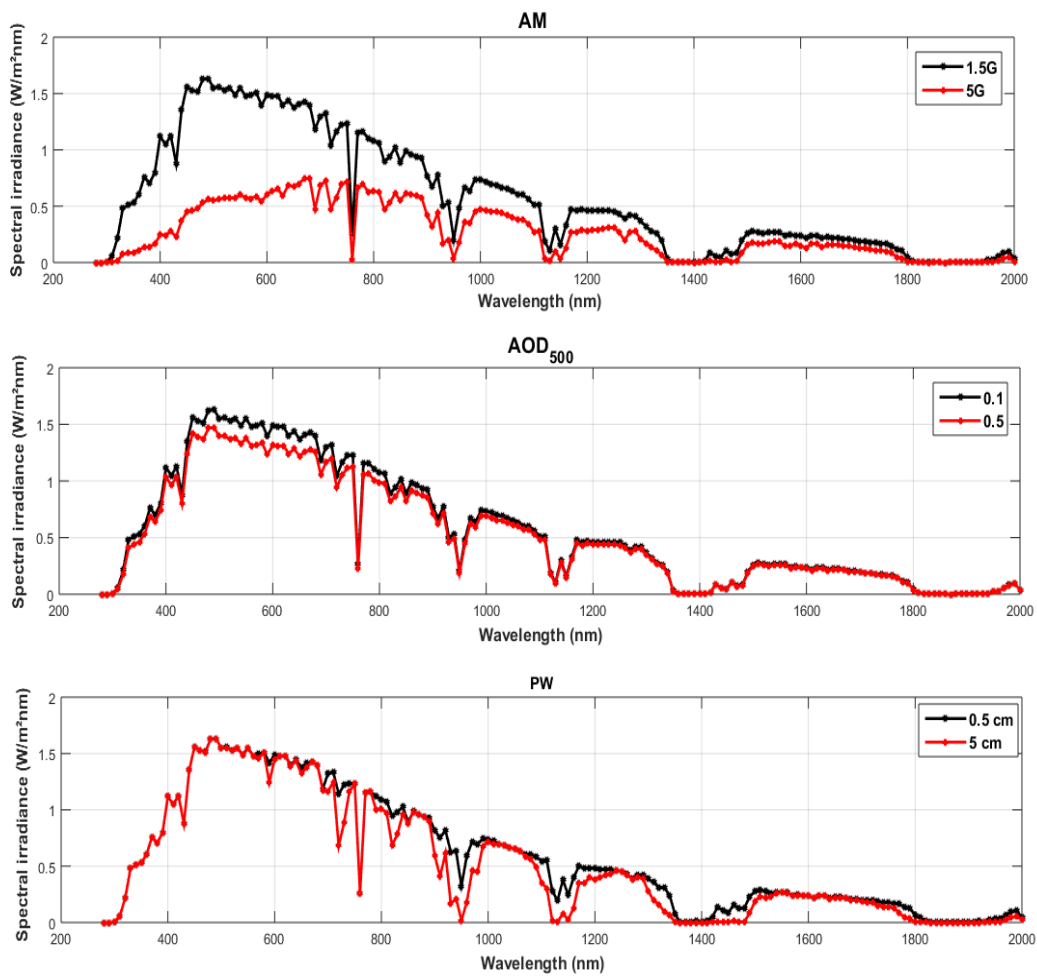

*Fig. 15. Top: Impact of air mass on the global spectral irradiance. The other parameters: AOD and PW are kept constant at 0.1 and 1 cm respectively. Middle: Impact of AOD at 500 nm on the global spectral irradiance. The other parameters: AM and PW are kept constant at 1.5 and 1 cm respectively. Bottom: Impact of PW on the global spectral irradiance. The other parameters: AM and AOD are kept constant at 1.5 and 0.1 respectively.*

## 4.1. Soiling

Fig. 16 shows the impact of soiling on the hemispherical transmittance of a One Diamant® low-iron glass coupon with a size of 4 cm x 4 cm in size and a thickness of 3 mm after 13 weeks of horizontal outdoor exposure in Jaén, Spain [103], a town with a high annual horizontal irradiation, 1790 kWh/m², and a Mediterranean-continental climate [65]. The PW and AOD values can be considered low-medium; nevertheless the second ones may seasonally reach peculiarly extreme values because of specific events, such as dust storms from the Sahara Desert or the presence of pollen in air from olive trees in the region [147]. As can be noticed, soiling causes an attenuation of the hemispherical transmittance of the coupon for all wavelengths, and its impact is higher on the blue region of the spectrum, with losses up to 45% at 300 nm.

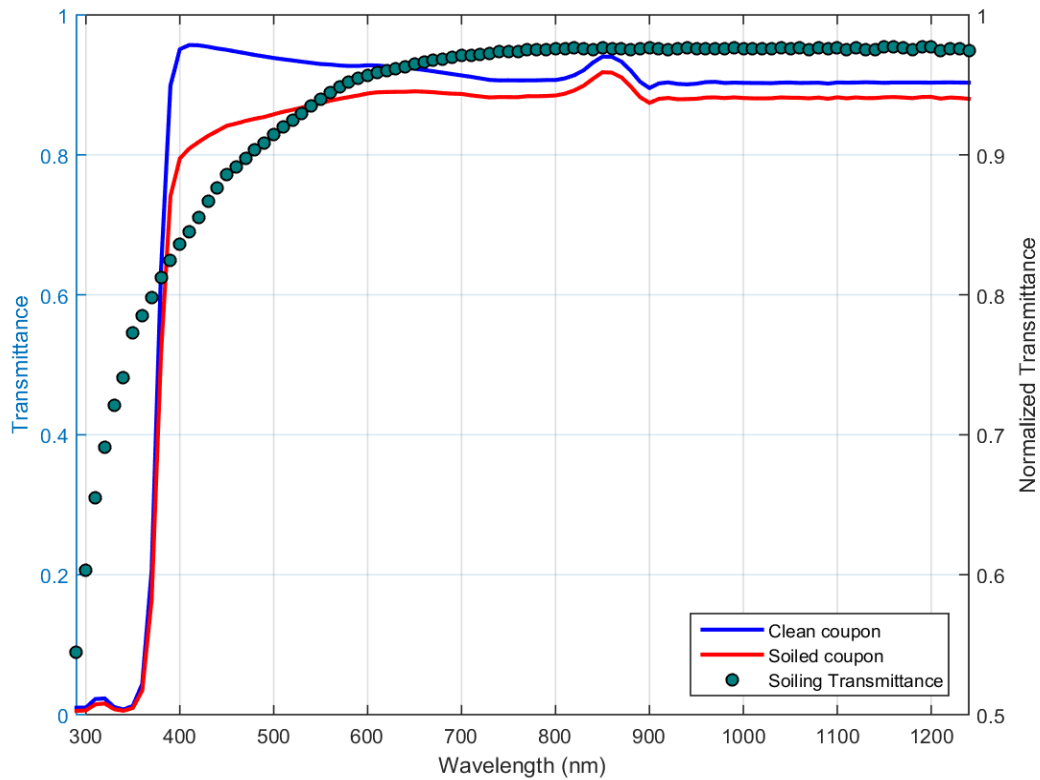

*Fig. 16. Left: Hemispherical transmittance of a reference clean coupon (blue) and hemispherical transmittance of the coupon after 24 weeks of exposure. Right: Normalized soiling transmittance obtained through Eq. 8.*

### 4.1.1. Impact of Air Mass (AM)

Fig. 17-left shows the impact of transmittance losses caused by soiling on the spectral behaviour as a function of AM for six different PV technologies. As can be seen, soiling produces spectral gains, (i.e. the current lowers less than the irradiance) and thus the value of the SDRatio is higher than one, for all the technologies considered with the exception of a-Si. This allows to state that during the hours with higher irradiance values, which are also those with lower AM values, the spectral gains originated by soiling contribute to the reduction of energy losses. Therefore, in locations with higher soiling accumulations, spectral gains can be relevant and the amount of energy losses can be minimized. Spectral gains greater than 1% are observed for m-Si, p-Si and CIGS technologies with low AM values in this specific location. On the other hand, a-Si technology presents spectral losses higher than 2% for AM values greater than 4 as a consequence of its narrow absorption band, which is mostly situated in the UV and visible regions of the spectrum where transmittance losses caused by soiling are maximum. However, it should be highlighted that AM values greater than 4 occur during the hours with lower solar irradiation, when energy production is smaller, limiting the significance of the losses.

### 4.1.2. Impact of Aerosol Optical Depth (AOD)

Fig. 17-middle shows the impact of transmittance losses caused by soiling on the spectral behaviour as a function of AOD of six different PV technologies. As in the previous case, soiling produces spectral gains for five of the six technologies under consideration, although unlike the previous one, the values of these gains remain almost constant for all AOD between 0 and 1. For AOD = 0.1, the value of spectral gains is 1.24% for m-Si. On the other hand, a-Si technology presents spectral losses around 1.5% within the full range of AOD values. The lesser impact of AOD on the spectra in comparison with AM accounts for the slight variations of SDRatio with this parameter for all PV technologies. Moreover, it can be seen that the trend of all of them is the same and this fact can be justified considering that the higher variations of the spectrum with AOD occur at wavelengths between 430 nm and 800 nm, where the normalised spectral response of all the technologies considered reaches values over 0.75.

### 4.1.3. Impact of Precipitable Water (PW)

Fig. 17-right shows the impact of transmittance losses caused by soiling on the spectral behaviour as a function of PW of six different PV technologies. As in the two previous cases, soiling produces spectral gains for all the technologies considered, excepting a-Si. In this case, a slightly increase in these gains, and a lightly decrease in losses for a-Si, with PW is observed, reaching maximum values when PW = 5 cm. These values are very similar to those mentioned before when considering the AOD impact. The same argument used for AOD can be used to justify the equal trend of all PV technologies.

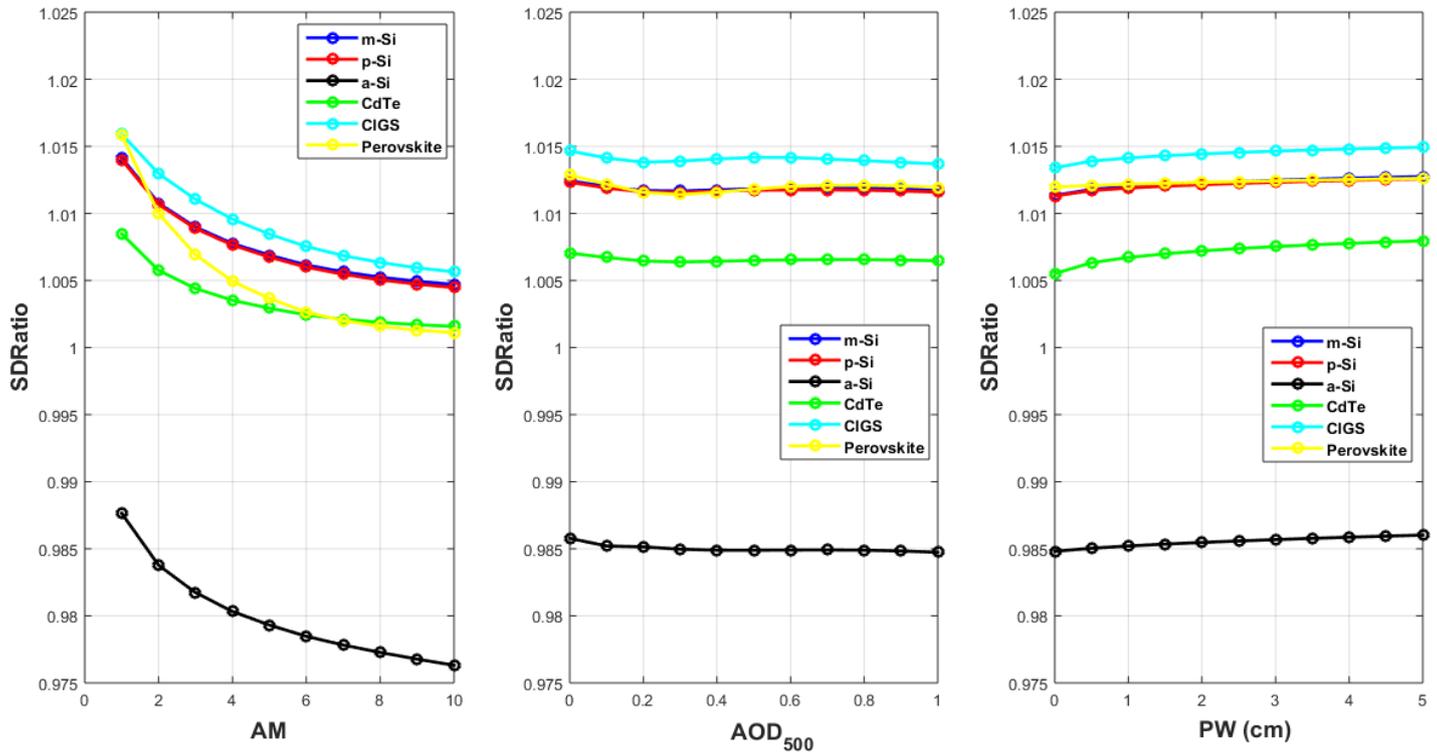

*Fig. 17. Spectral degradation ratio due to soiling as a function of the individual variation of AM (Left), AOD (Middle) and PW (Right) for six different PV technologies. For each case, the two others parameters are kept constant (AM 1.5, AOD = 0.1 and PW = 1 cm).*

### 4.2. Discoloration

Fig. 18 shows the impact of discoloration on the hemispherical transmittance of the EVA encapsulant sheet of a m-Si PV module after 25 years of outdoor exposure in Yeosu, South Korea, as reported in [125].

As can be observed, at shorter wavelengths, between 300 nm and 360 nm, EVA discoloration causes an increment of the transmittance, which the authors of the study associate to the loss of UV absorbers in the encapsulant; whereas for wavelengths higher than 400 nm, discoloration produces transmittance losses with minimum and maximum values of 10% and 50% at 840 nm and 400 nm respectively.

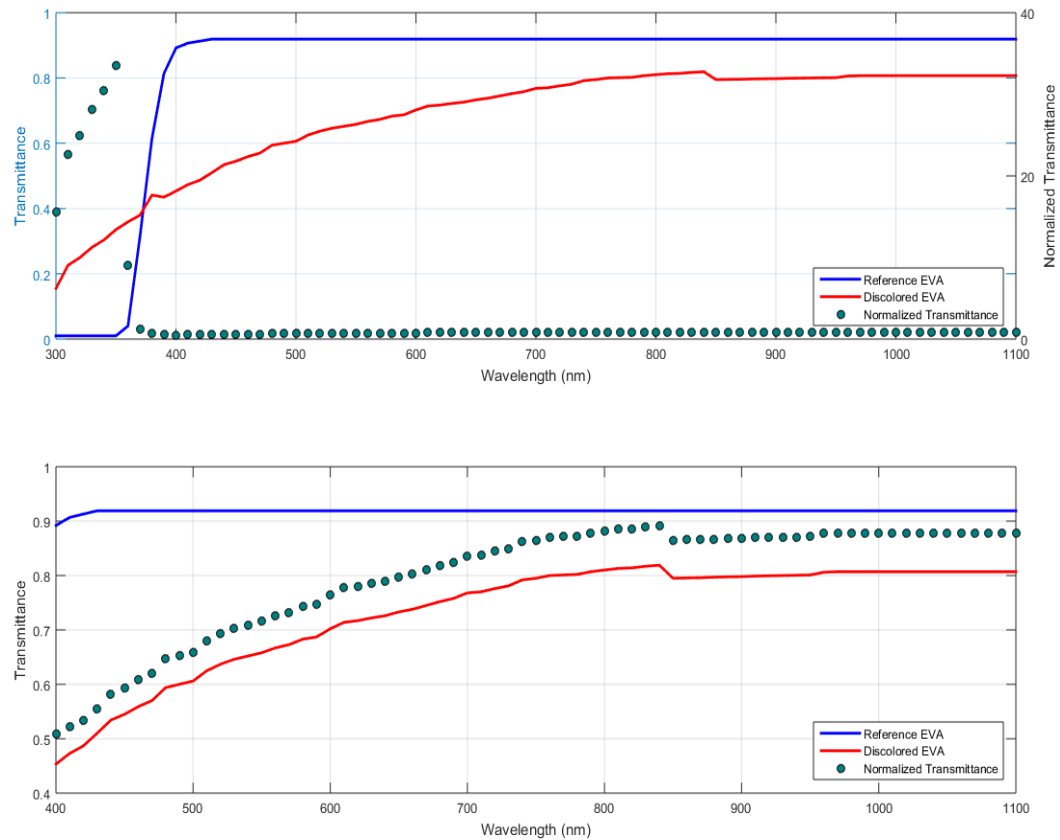

*Fig. 18. Top: Transmittance of the EVA sheet of a PV module before and after discoloration (Left) and Normalized Transmittance (Right). Bottom: Transmittance in visible and near - infrared regions of the spectrum*

### 4.2.1. Impact of Air Mass (AM)

Fig. 19-left shows the impact of transmittance losses caused by EVA sheet discoloration on the spectral performance as a function of AM of six different PV technologies. As can be seen, discoloration produces spectral losses for all the PV technologies under consideration. These losses are greater at low AM values; for AM = 1, spectral losses between 6% (CIGS) and 60% (Perovskite) are obtained. A significant reduction of spectral losses, which corresponds to an increase of the SDRatio, is appreciated with a rise of AM. This can be explained by the combination of the transmittance profile (Fig. 18) and the reduction of the spectral irradiance with AM. Indeed, the magnitude of the reduction in intensity that PV devices can generate is significantly smaller that the magnitude of the spectral irradiance decrease in the blue region of the spectrum.

It should be highlighted that CIGS devices show limited spectral losses. These losses reach their maximum (6%) at AM = 1, and are negligible for AM values greater than 5. This can be explained by the reduced spectral response in the UV part of the spectrum, which is the region where the transmittance of the discoloured EVA sheet is lower than the transmittance of the reference EVA sheet. This fact usually occurs on the centre of the cells due to the complete loss of UV absorbers.

### 4.2.2. Impact of Aerosol Optical Depth (AOD)

Fig. 19-middle shows the impact of transmittance losses caused by EVA sheet discoloration on the spectral performance as a function of AOD of six different PV technologies. As in the previous case, discoloration produces spectral losses for all the technologies considered. However, in this case, spectral losses remain almost constant with AOD values. In the case of Perovskite PV devices, maximum spectral losses occur when no aerosols are present (54%) and minimum losses when AOD is 0.3 (51%).

### 4.2.3. Impact of Precipitable Water (PW)

Fig. 19-right shows the impact of transmittance losses caused by EVA sheet discoloration on the spectral performance as a function of PW of six different PV technologies. The results are very similar to those obtained by the analysis of the impact of AOD. In this case, a slightly increase of spectral losses with PW is observed, reaching maximum values when PW = 5 cm.

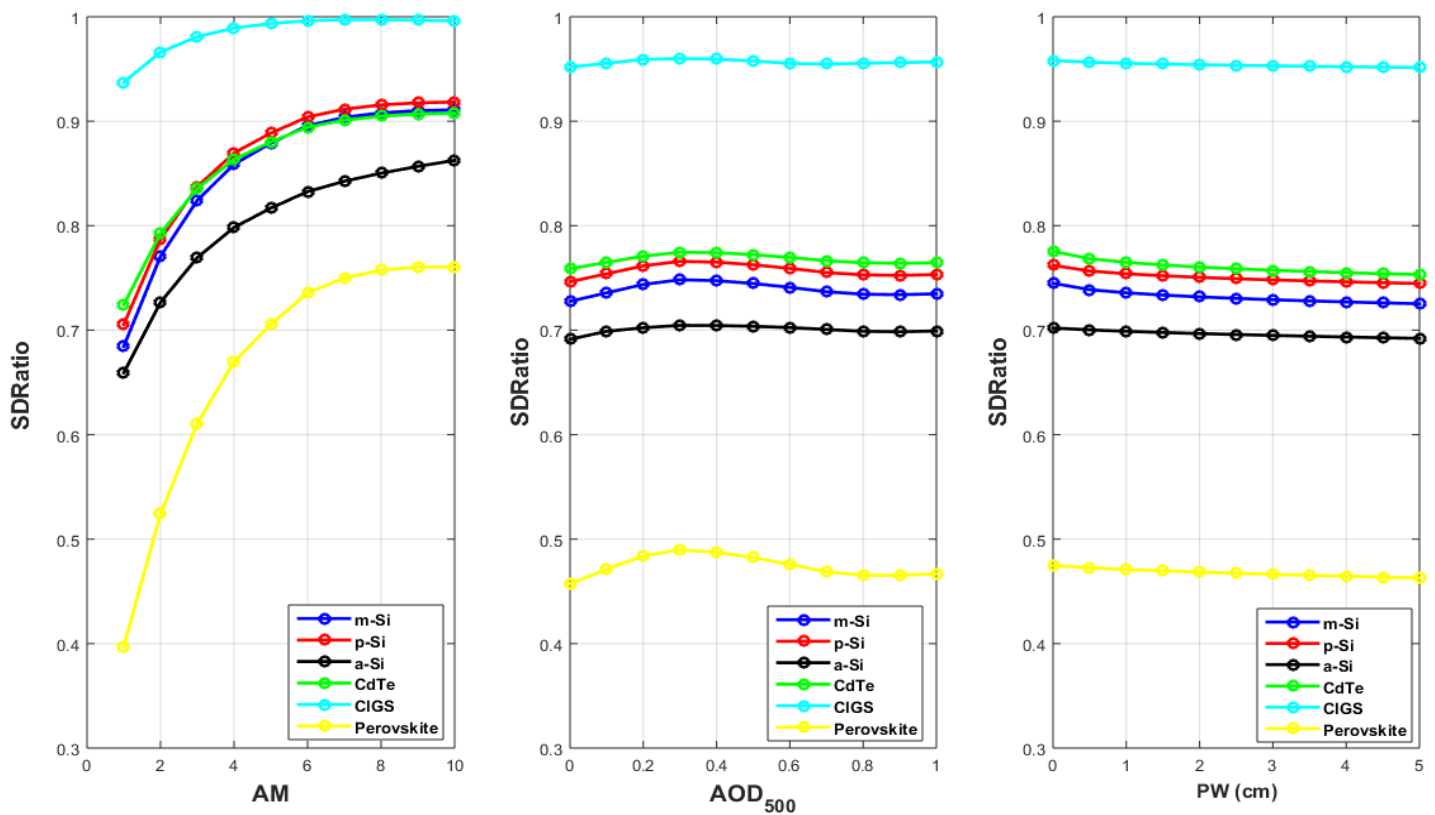

*Fig. 19. Spectral degradation ratio due to discoloration as a function of the individual variation of AM (Left), AOD (Middle) and PW (Right) for six different PV technologies. For each case, the two others parameters are kept constant (AM 1.5, AOD = 0.1 and PW = 1 cm).*

## 5.  Conclusions

PV modules performance presents a high dependence on the spectral response. In this review, the optical degradation factors that substantially affect both the transmittance of the PV encapsulant and the spectral response, are presented. A decrease in transmittance or in spectral response leads to a reduction of the short-circuit current density of the modules, thus affecting the electrical performance. In addition, the effects of soiling are assessed in depth with a table that includes the most relevant studies performed during the last years. The aim of this table is to provide the PV community with an overview of this issue. As it can be seen in this work, many studies have been performed, especially in the last 10 years, highlighting the interest of the topic within the field of photovoltaic solar energy. The vast majority of these studies presents results related to the effects on the spectral transmittance of the encapsulant of m-Si and p-Si commercial modules. These two PV technologies are characterized by the frequent use of plastic-materials as top-sheet encapsulant; in contrast with other PV technologies, such as CdTe modules that have a glass encapsulant. This fact plays a key role when evaluating the optical degradation. This work intends to promote investigations that analyse how optical degradation affects different PV technologies, such as CdTe and a-Si, with different encapsulant materials.

The analysis of the spectral impact of soiling and discoloration on different PV technologies is also presented in this paper. The variations of the incident spectral irradiance due to several atmospheric parameters (AM, AOD and PW) have been considered. The simulations show that soiling produces spectral gains for all the technologies considered, with the exception of a-Si. Maximum values around 1% are obtained for a location with intermediate levels of soiling accumulations at AM values below 2.5. These spectral gains originated by soiling can contribute to the reduction of energy losses, particularly in the hours with high irradiance. Furthermore, in locations with higher soiling accumulations, spectral gains can be relevant and the amount of energy losses is minimized. Moreover, it has been proven that AOD and PW have negligible incidence on the results. On the other hand, it has been demonstrated that discoloration causes significant spectral losses for all PV technologies, with the exception of CIGS. The values of these losses present an important dependence on AM, as they experiment a gradual decrease with this parameter (in the instance of CdTe technology, maximum and minimum values are 28% with AM 1.0 and 9% with AM 10 respectively). As in the case of soiling, AOD and PW do not show changes of great importance upon the results.

In this work, numerous studies investigating the effects of different optical degradation mechanisms on PV devices have been addressed, however only a few of them show graphical data that allow the spectral characterization of degradation. The detailed analysis of the spectral impact of soiling and discoloration presented here, which is based in actual and measured data, expect the promotion of more studies in this issue.

## Acknowledgements

Álvaro F. Solas, the corresponding author of this work, is supported by the Spanish ministry of Science, Innovation and Universities under the program "Ayudas para la formación de profesorado universitario (FPU), 2018 (Ref. FPU18/01460)". Part of this work was funded through the European Union's Horizon 2020 research and innovation programme under the NoSoilPV project (Marie Skłodowska-Curie grant agreement No. 793120). This study is partially based upon work from COST Action PEARL PV (CA16235), supported by COST (European Cooperation in Science and Technology). COST (European Cooperation in Science and Technology) is a funding agency for research and innovation networks. Our Actions help connect research initiatives across Europe and enable scientists to grow their ideas by sharing them with their peers. This boosts their research, career and innovation, see [www.cost.eu](www.cost.eu).

Figure 10 is reprinted from Han H, Dong X, Li B, Yan H, Verlinden PJ, Liu J, et al. Degradation analysis of crystalline silicon photovoltaic modules exposed over 30 years in hot-humid climate in China. Sol Energy 2018;170:510–9. https://doi.org/10.1016/j.solener.2018.05.027 wirh permission from Elsevier (License number: 4872950774748). Figure 11 is reprinted from Ndiaye A, Charki A, Kobi A, Kébé CMF, Ndiaye PA, Sambou V. Degradations of silicon photovoltaic modules: A literature review. Sol Energy 2013;96:140–51. https://doi.org/10.1016/J.SOLENER.2013.07.005 with permisssion from Elsevier (License number: 4872941308632) .